# NEW UNIVERSALITY CLASS FOR STEP BUNCHING IN THE "$C^+$ - $C^-$" MODEL

B. Ranguelov[1], V. Tonchev[1,*], H.Omi[2], A.Pimpinelli[3]
[1]Institute of Physical Chemistry, Bulgarian Academy of Sciences, 1113 Sofia, Bulgaria
[2]NTT Basic Research Laboratories, Atsugi, Kanagawa 243-0198, Japan
[3]LASMEA, CNRS/Université Blaise-Pascal, Aubiére, France

* Corresponding author: tonchev@ipc.bas.bg

**Abstract**. We formulate a new (1+1)D step model of potentially unstable vicinal growth that we call "$C^+$ - $C^-$" model and study the step bunching process in it. The basic assumption is that the equilibrium adatom concentrations on both sides of the step are different and this may cause destabilization of the regular step train. We deduce equations of step motion and numerically integrate them to obtain the step positions on a discrete time set. New dynamic phenomena are observed during the bunching process: (i) parts of the crystal surface undergo temporary evaporation in the course of coalescence of two bunches, the larger being the one that 'goes back', (ii) the minimal interstep distance appears in the beginning rather than in the middle of the bunch. We speculate that the latter may serve as a diagnostic criterion for a new universality class constructed from the time- and size-scaling exponents describing the bunching process in the diffusion-limited regime. In particular the size-scaling of the minimal interestep distance $l_{\min}$ for first time is obtained as $l_{\min} \sim N^{-1/(n+1)}$, where $N$ is the number of steps in the bunch and $n$ is the exponent in the step-step repulsions law $U{\sim}1/d^n$ for two steps placed a distance $d$ apart.

**Introduction.** The interest towards the introduction of novel models of surface instabilities was recently stimulated by experiments where simultaneous step bunching and step meandering was observed on growing vicinals of Cu[1,2] and Si[3]. One important characteristic of the new type of bunching is that the size-scaling exponent of the average interstep distance is 0.29±0.05[1,2] while in most cases of experimental and theoretical studies it was found as 2/3[4]. In this paper we introduce the so called "$C^+$ - $C^-$" model and obtain both size- and time-scaling relations for the step bunches formed in the long times limit. Our results could be considered as a challenge for further theoretical and experimental efforts to clarify and extend the existing universality classes scheme of Pimpinelli-Tonchev-Videcoq-Vladimirova (PTVV) [5,4].

**The Model.** The "$C^+$ - $C^-$" model is an extended BCF[6] model. One considers (1+1)D vicinal crystal surface with steps descending in the $+x$ direction. First one obtains the concentration of the adatoms $C(x)$ on a single terrace assuming that the concentration fields on the neighboring terraces are not conjugated otherwise than through the step as a source/sink of adatoms. The additional assumption that the concentration field adjusts immediately to the step positions permits to consider the steps as motionless. Thus the stationary diffusion equation is rather simple:

$$D_s \frac{\partial^2 C}{\partial x^2} + F = 0 \tag{1}$$

where $F$ is the incoming flux, $D_s$ is the surface diffusion coefficient of the adatoms and the desorption is neglected. This equation is integrated twice leading to:

$$C(x) = -\frac{Fx^2}{2D_s} + C_1 x + C_2 \tag{2}$$

The integration constants $C_1$ and $C_2$ are obtained using boundary conditions on both steps that enclose the terrace:

$$D_s \left.\frac{\partial C}{\partial x}\right|_{x=x_i} = K\left[C(x_i) - C_e^+(x_i)\right] \quad \text{and} \quad D_s \left.\frac{\partial C}{\partial x}\right|_{x=x_{i+1}} = -K\left[C(x_{i+1}) - C_e^-(x_{i+1})\right] \tag{3}$$

where $K$ is the kinetic coefficient (equal from both sides of the step) and the equilibrium concentrations on both sides of the step $C_E$ (E=$L$, $R$) are affected by step-step repulsions with energy $U = A/d_0^n$ ($A$ is the magnitude of the repulsion and $d_0$ is the interstep distance) and given as :

$$C_e^+(x_i) = C_R\left[1 + l_0^{n+1}\left(\frac{1}{(x_{i+1} - x_i)^{n+1}} - \frac{1}{(x_i - x_{i-1})^{n+1}}\right)\right] \tag{4}$$

$$C_e^-(x_{i+1}) = C_L\left[1 + l_0^{n+1}\left(\frac{1}{(x_{i+2} - x_{i+1})^{n+1}} - \frac{1}{(x_{i+1} - x_i)^{n+1}}\right)\right] \tag{5}$$

with $l_0 = (nA\Omega/kT)^{\frac{1}{n+1}}$ , $\Omega$ is the crystal surface area per atom. Usually $n$ takes the *canonical* value[7] $n$=2 but the use of the more general notation $n$ permits to study systematically the scaling behavior of the model. The basic assumption of the model is that the equilibrium concentrations on both sides of the step are different:

$$C_L \neq C_R \tag{6}$$

This difference is expected to destabilize the growing surface.

The model was obtained[15] as an effective step model from a two-particle model[8,9]. The problem with the physical justification of the model remains for further studies but what should be pointed out is that modification of the equilibrium concentrations is also possible due to different factors as step curvature [10], step transparency [11] and electromigration of adatoms [12].

The concentration profile on the terrace is obtained as:

$$C(x) = \frac{F}{2D_s}\left[(x_{i+1} - x)(x - x_i) + d(x_{i+1} - x_i)\right] + \\ + \frac{(x - x_i)\left[C_e^-(x_{i+1}) - C_e^+(x_i)\right] + d\left[C_e^-(x_{i+1}) + C_e^+(x_i)\right] + C_e^+(x_i)(x_{i+1} - x_i)}{2d + x_{i+1} - x_i} \tag{7}$$

where the notation $d \equiv D_s / K$ is used, $d$ usually called *kinetic length*[10]. Further one obtains the step velocity as a sum of two contributions – the mass fluxes from the two terraces neighboring a step:

$$\frac{1}{D_s\Omega}\frac{dx_i}{dt} = \frac{\partial C^i(x_i)}{\partial x} - \frac{\partial C^{i-1}(x_i)}{\partial x} \tag{8}$$

arriving at:

$$V_i = \frac{dx_i}{dt} = \Omega D_s\left(-\frac{\partial C}{\partial x}\bigg|_{x=x_i}^{i-1} + \frac{\partial C}{\partial x}\bigg|_{x=x_i}^{i}\right) = \Omega D_s\left(C_1^i - C_1^{i-1}\right) =$$

$$F\Omega\left(\frac{x_{i+1}-x_{i-1}}{2}\right) + \frac{\Omega D_s}{2d+\Delta x_{i+1}}\left\{C_L\left[1+l_0^{n+1}\left(\frac{1}{\Delta x_{i+2}^{n+1}} - \frac{1}{\Delta x_{i+1}^{n+1}}\right)\right] - C_R\left[1+l_0^{n+1}\left(\frac{1}{\Delta x_{i+1}^{n+1}} - \frac{1}{\Delta x_{i}^{n+1}}\right)\right]\right\} -$$

$$-\frac{\Omega D_s}{2d+\Delta x_{i}}\left\{C_L\left[1+l_0^{n+1}\left(\frac{1}{\Delta x_{i+1}^{n+1}} - \frac{1}{\Delta x_{i}^{n+1}}\right)\right] - C_R\left[1+l_0^{n+1}\left(\frac{1}{\Delta x_{i}^{n+1}} - \frac{1}{\Delta x_{i-1}^{n+1}}\right)\right]\right\} \tag{9}$$

As seen the velocity of the $i$-th step is obtained as function of model parameters as the flux $F$, the surface diffusion constant $D_s$, kinetic coefficient $K$, magnitude of the interstep repulsions $A$ and differences of the type $\Delta x_i = x_i - x_{i-1}$ representing the widths of the corresponding terraces (the terrace between $i$-th and the $i$-1-st step in this example). The above equation is transformed into dimensionless form with four dimensionless parameters:

$$\frac{d}{l} = \frac{1}{l}\frac{D_s}{K};\ \alpha \equiv \frac{C_L}{C_R};\ \frac{D_s C_R}{Fl^2};\ \frac{l_0}{l} = \frac{1}{l}\left(\frac{nA\Omega}{kT}\right)^{\frac{1}{n+1}}$$

where $l$ is the initial vicinal distance.

**Linear Stability Analysis.** We carry out a simple linear stability analysis on the obtained equation for step velocity, Eq.(9), perturbing only the $i$-th step in the moving equidistant step train and thus changing the widths of the two neighboring terraces:

$$\Delta x_i \to l + \delta l,\ \Delta x_{i+1} \to l - \delta l$$

where $l$ is the initial (vicinal) interstep distance. Then, what determines the (in)stability of the step train is the sign of $\delta V_i(\delta l)$ - when it is (positive)negative the step train is (un)stable. After some algebraic operations one obtains the instability condition in the form:

$$\left(\frac{2}{3(n+1)}\right)^{\frac{1}{n+1}} > \left(\frac{l_0}{l}\right)\left[\frac{\left(2\frac{d}{l}+1\right)}{\left(\frac{\alpha-1}{\alpha+1}\right)}\right]^{\frac{1}{n+1}} \equiv S \tag{10}$$

As seen, the stability of the regular step train is not function of the ratio $D_s C_R / Fl^2$. In a subsequent study we will show that this parameter also does not enter in the scaling relations describing the late stages of the bunching process and what determines the system's behavior in this stage is only the combination of parameters denoted with *S*. The expression in the l.h.s. of the above inequality could be roughly approximated with (0.555.n+0.027) in the range n=1÷7 and could eventually be used as a definition of whether an interstep distance is a bunch one. In fact, the analysis of the linear stability of the whole step train shows that the instability condition is simpler, $C_L > C_R$, but we leave the discussion on this issue for a further publication.

**Results.** We integrate the equations of step motion using fourth order Runge-Kutta procedure[13] (note that in this paper we restrict our study to the diffusion-limited regime of growth, $d<<1$) and obtain step positions on a discrete time set. These step trajectories are shown on fig.1a. Two new dynamical phenomena are illustrated on fig.1b: (i) parts of the crystal surface undergo temporary evaporation and this leads to coalescence of two bunches, the larger being the one that 'goes back', (ii) the minimal interstep distance appears in the beginning rather than in the middle of the bunch. Thus bunches have a maximal slope where steps join them after leaving the bunch from behind and crossing the terraces. The dependence of the minimal bunch distance $l_{\min}$ on the number of steps *N* in the bunch is shown of fig.2 for different values of *n*, the exponent in the step-step repulsions law. The obtained dependence $l_{\min} \sim N^{-1/(n+1)}$ differs from the one for the classical case of bunching due to electromigration or Ehrlich-Schwoebel effect[4] which is $l_{\min} \sim N^{-2/(n+1)}$. In the latter case the minimal interstep distance appears in the middle of the bunch and the scaling behavior of the first distance is $l_1 \sim N^{-1/(n+1)}$. Thus the "$C^+$ - $C^-$" model is both different from and similar to the classical case. In fig.3 is plotted the bunch width $L_b$ vs. bunch size *N* again for different values of *n*. The resulting dependence is $L_b \sim N^{n/(n+1)}$. Time dependencies of the bunch size *N* and the bunch width $L_b$ are plotted in fig.4. As seen, the time-scaling exponent $\beta$=1/2 of the bunch size is not dependent on *n* and this behavior is analogous to the $\rho$=-1 universality class in the PTVV classification scheme [5] and the time-scaling exponent of the bunch width $L_b$ is 1/z = n/[2(n+1)], while in PTVV it is $1/z = (n-\rho)/2(n+1-2\rho)$ and they are the same only for $\rho = 0$. The results from the complete scaling analysis of the model are collected in Table 1 together with data from other models.

**Conclusion.** The model is believed to recover the behavior of a class of step bunching phenomena characterized qualitatively with the appearance of the minimal interstep distance in the beginning of the bunch. Quantitatively, the size- and time-scaling exponents are different from those, predicted by PTVV [5,4] and fall into a new universality class. Further theoretical and experimental efforts could show the proper place of the model studied here. At least, it will remain as one of the reference models for the further extension of the universality classes for step bunching phenomena.

**Acknowledgement.** Most calculations in this study were done on a computer cluster built with the financial support of grants F-1413/2004 and BM-9/2006 from the Bulgarian National Science Fund and VIRT/NANOPHEN-FP6-2004-ACC-SSA-2. VT acknowledges hospitality and very stimulating working conditions at the NTT Basic Research Laboratories, Morinosato Wakamiya, Japan.

Table 1. Universality classes obtained from the PTVV[5] classification scheme and compared to the new one generated by the "$C^+$ - $C^-$"- model. Shaded are some exponents that are invariant along the columns or rows.

| | $C^+$ - $C^-$ | Universality Classes PTVV[5] | | |
|---|---|---|---|---|
| **$1/\alpha$: $L_b$~$H^{1/\alpha}$** | n/(n+1) | (n-ρ)/(2+n-ρ) | | |
| **$\beta=\alpha/z$: H~$t^\beta$** | 1/2 | (2+n-ρ)/2(n+1-2ρ) | | |
| **$1/z$: $L_b$ ~ $t^{1/z}$** | n/2(n+1) | (n-ρ)/2(n+1-2ρ) | | |
| **$\delta$: $l_b$ ~ $t^{-\delta}$** | 1/2(n+1) | 1/(n+1-2ρ) | | |
| **$\gamma$: $l_b$ ~ $N^\gamma$** | 1/(n+1) | 2/(2+n-ρ) | | |
| | | **ρ = -2** | **ρ = -1** | **ρ = 0** |
| | **n=1** | | | |
| **1/α** | 1/2 | 3/5 | 1/2 | 1/3 |
| **β** | 1/2 | 5/12 | 1/2 | 3/4 |
| **1/z** | 1/4 | 1/4 | 1/4 | 1/4 |
| **δ** | 1/4 | 1/6 | 1/4 | 1/2 |
| **γ** | 1/2 | 2/5 | 1/2 | 2/3 |
| | **n=2** | | | |
| **1/α** | 2/3 | 2/3 | 3/5 | 1/2 |
| **β** | 1/2 | 3/7 | 1/2 | 2/3 |
| **1/z** | 1/3 | 2/7 | 3/10 | 1/3 |
| **δ** | 1/6 | 1/7 | 1/5 | 1/3 |
| **γ** | 1/3 | 1/3 | 2/5 | 1/2 |

Figure 1. Step trajectories (a) and surface slope (b) obtained for the "$C^+$ - $C^-$" model. Note on (a) the transition from a regime of bunching without steps on the terraces towards a regime where single steps cross regularly the terraces. This regime was first obtained in the minimal models [14] and later in the "$C^+$ - $C^-$"model. Note also in (b) that the surface is steepest in the beginning of the bunch (this is the side where steps join the bunch after crossing the terrace).

Figure 1a.

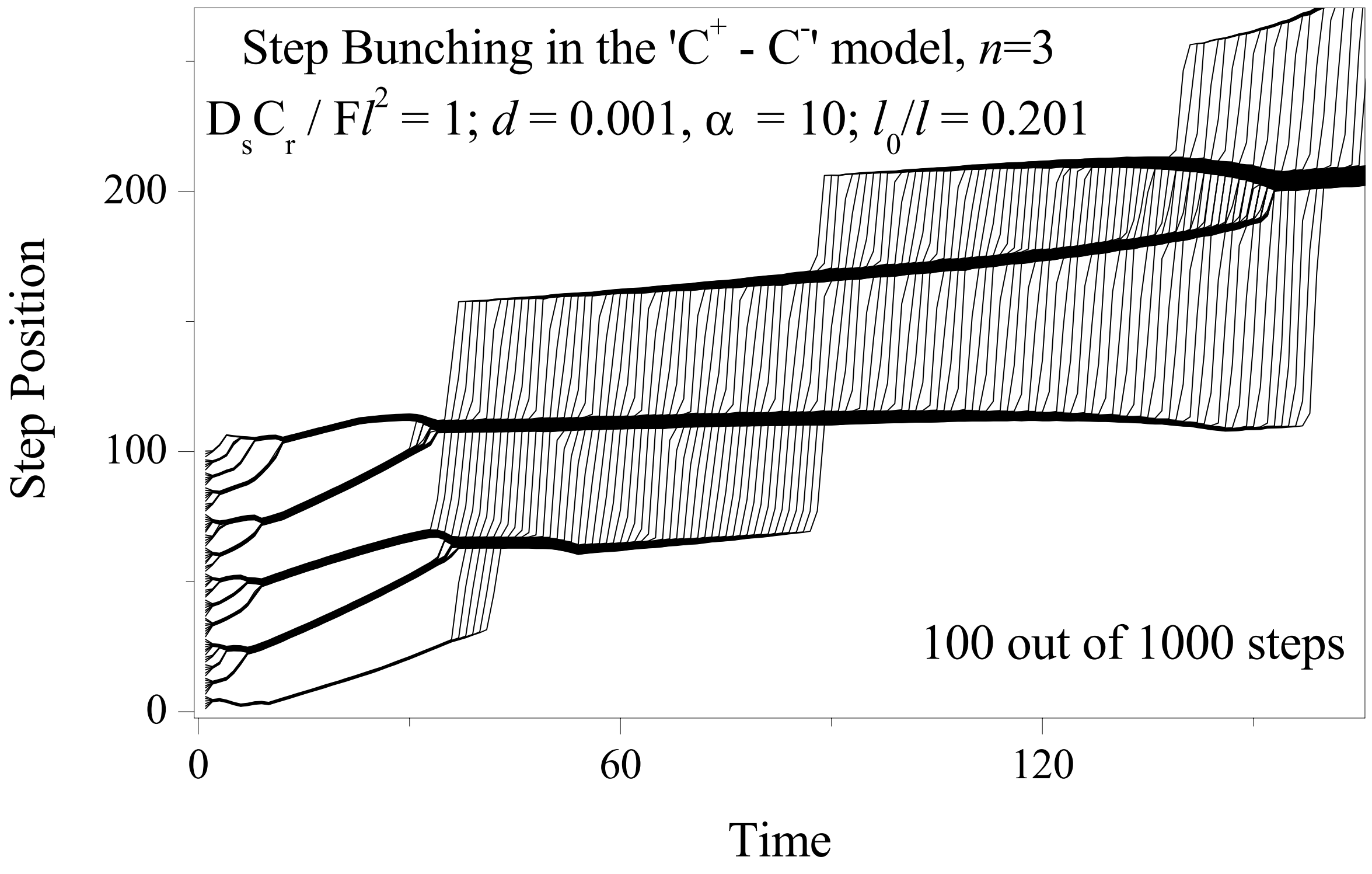

Figure 1b.

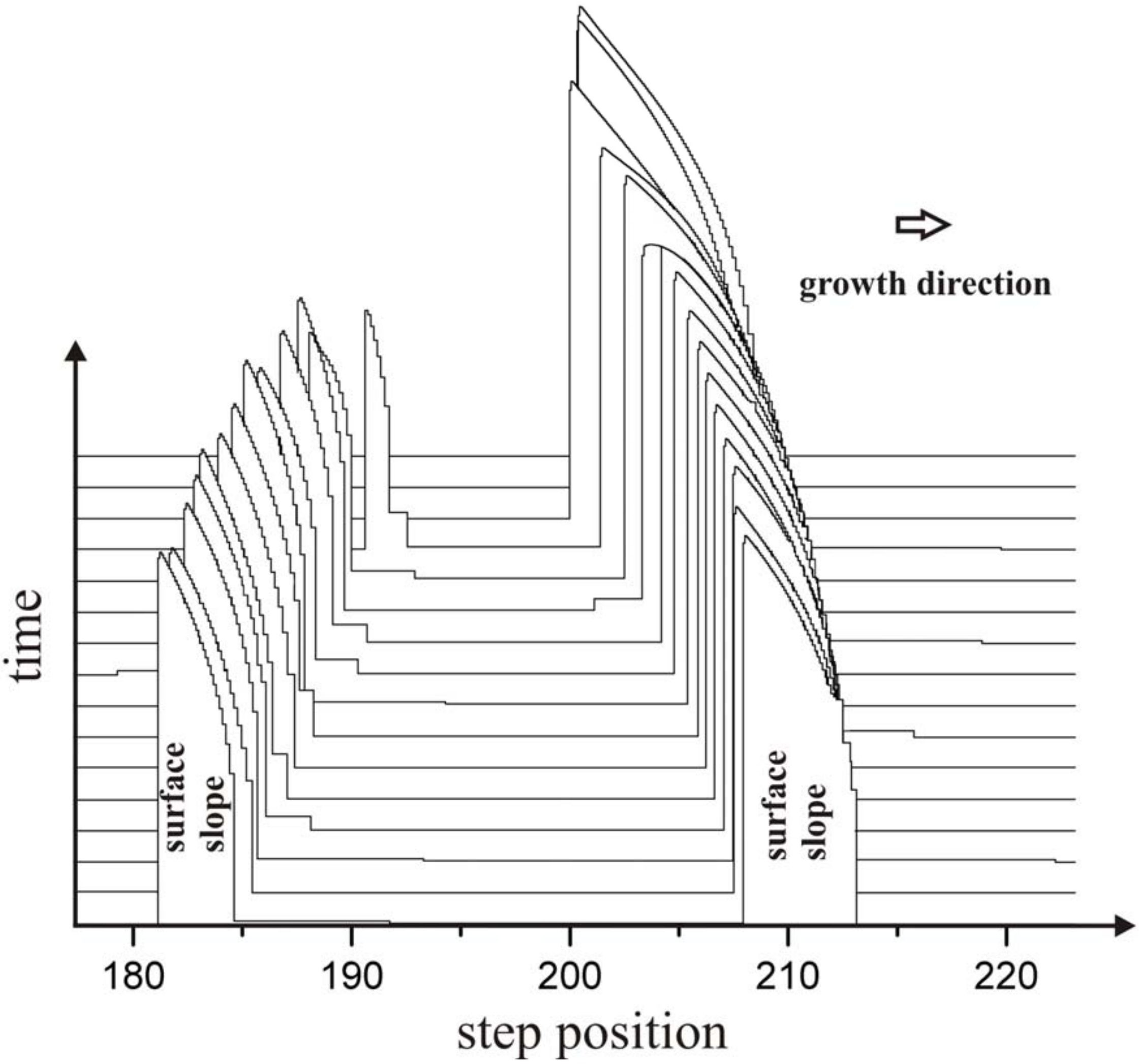

Figure 2. Minimal interstep distance in the bunch $l_{\min}$ as function of the number of steps in the bunch $N$. For different values of $n$ the values of the step-step repulsion parameter $l_0$ are also different.

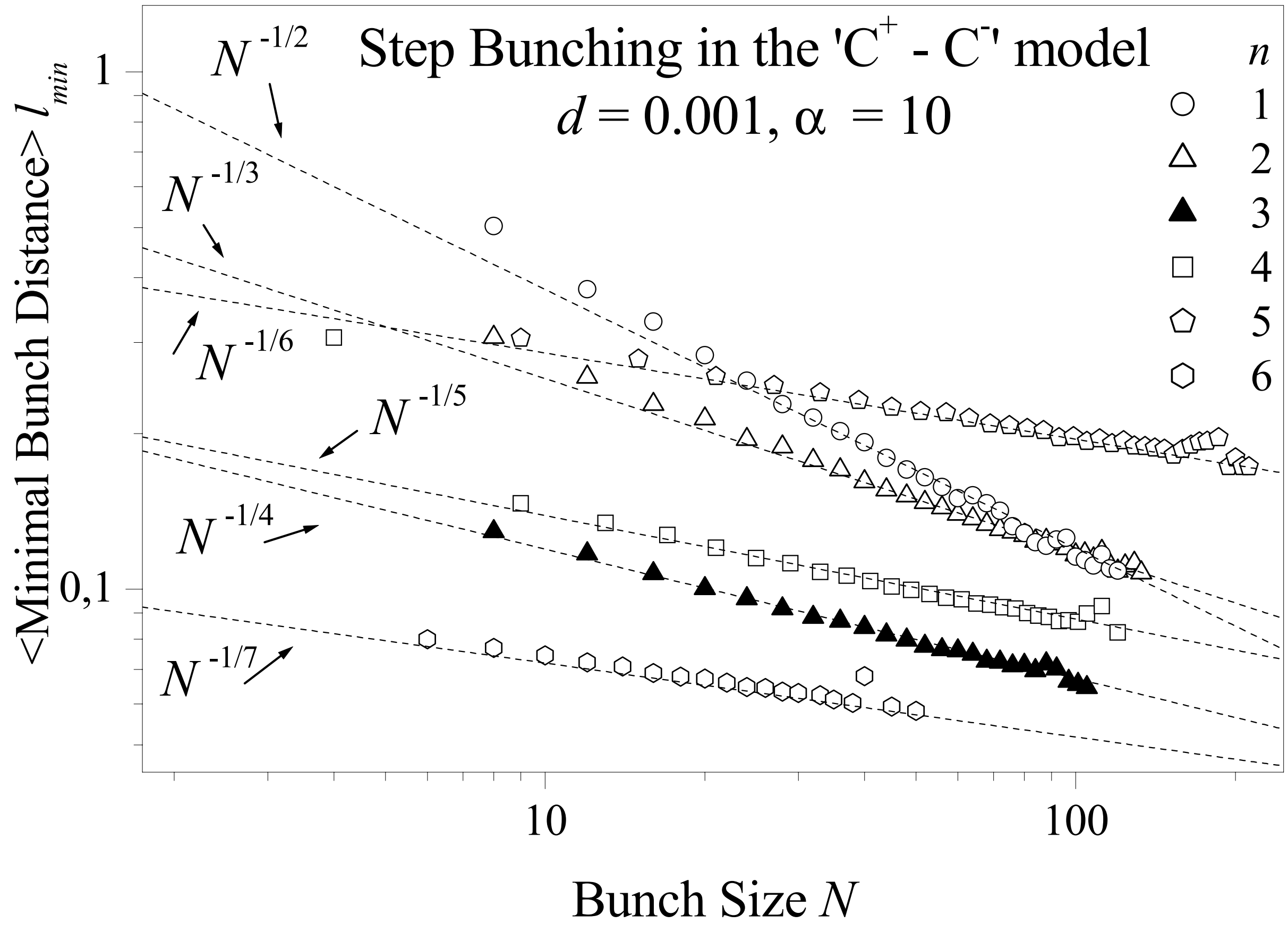

Figure 3. Bunch width vs. bunch size dependence. Only part of the data sets used in Fig.2 is shown.

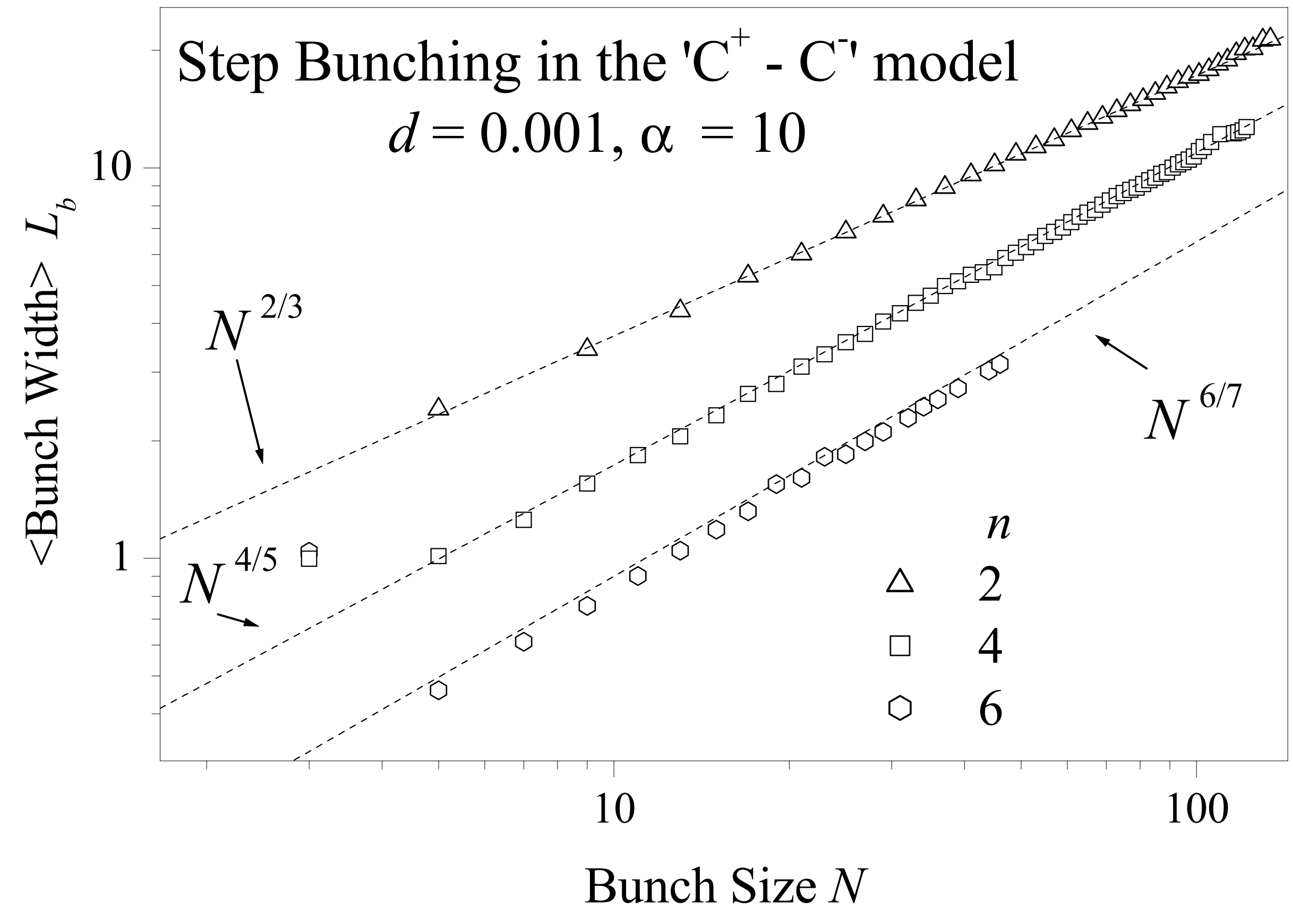

Figure 4. Time dependence of the bunch size and bunch width. The same data sets as on Fig.3 are shown. Note that the data for bunch size from different data sets almost overlap.

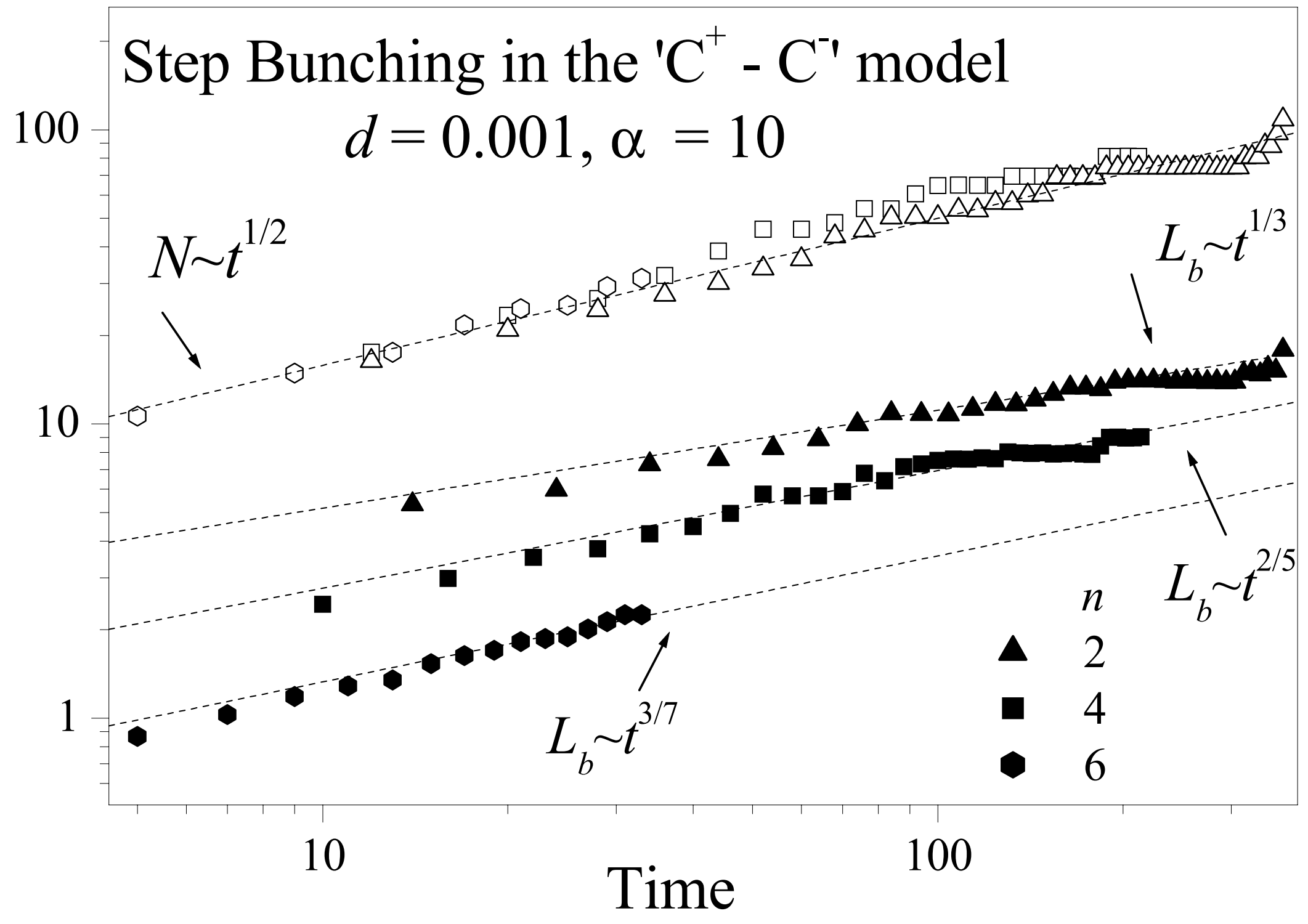